\documentclass[%
reprint,
twocolumn,
superscriptaddress,
showpacs,preprintnumbers,
nofootinbib,
aps,
prl,
]{revtex4-1}

\usepackage{amsmath}
\usepackage{amssymb}
\usepackage{mathtools}

\usepackage{hyperref}

\usepackage{graphicx}

\usepackage[dvips]{color}

\usepackage{ulem}

\usepackage{comment}

\graphicspath{
{./}
{./illustrations/}
}

\def\al{\alpha} 
\def\be{\beta} 

\def\de{\delta}
\def\ep{\epsilon}
\def\ze{\zeta}

\def\th{\theta}

\def\la{\lambda}

\def\ta{\tau}

\def\om{\omega}
\def\De{\Delta}

\def\pa{\partial}
\def\na{\nabla}

\newcommand{\ben}{\begin{equation}}
\newcommand{\een}{\end{equation}}
\newcommand{\bea}{\begin{eqnarray}}
\newcommand{\eea}{\end{eqnarray}}
\newcommand{\ba}{\begin{array}}
\newcommand{\ea}{\end{array}}
\newcommand{\bit}{\begin{itemize}}
\newcommand{\eit}{\end{itemize}}
\newcommand{\vev}[1]{\left\langle#1\right\rangle}

\newcommand{\half}{\frac12}

\newcommand{\bk}{\textbf{k}}
\newcommand{\bq}{\textbf{q}}

\newcommand{\bx}{\textbf{x}}

\newcommand{\dbar}[2]{\frac{d^{#1}{#2}}{(2\pi)^{#1}}}
\newcommand{\debar}[2]{{(2\pi)^{#1}}\de({#2})}

\newcommand{\cs}{c_\text{s}} % Sound speed
\newcommand{\vw}{v_\text{w}} % Wall velocity
\newcommand{\fluidV}{\overline{U}_\text{f}}  % Mean velocity
\newcommand{\Rbc}{R_*} % Mean droplet radius at collision
\newcommand{\Hc}{H_*} % Hubble rate at transition
\newcommand{\tLife}{\tau_\text{v}}
\newcommand{\Rfluid}{L_\text{f}}

\newcommand{\Pspec}[1]{{\mathcal P}_{#1}}
\newcommand{\SpecDen}[1]{P_{#1}}
\newcommand{\SpecDenGW}{\tilde P_{\text{GW}}}

\newcommand{\OmGW}{\Omega_\text{GW}}

\newcommand\vel[1]{v_{#1}}

\newcommand{\uetcTen}{\Pi^2}

\newcommand{\vip}{v_\text{ip}}
\newcommand{\vmax}{v_\text{max}}
\newcommand{\tInit}{t_\text{i}}
\newcommand{\TInit}{T_\text{i}}

\newcommand{\omtil}{\tilde{\omega}}
\newcommand{\qtil}{\tilde{q}}
\newcommand{\bqtil}{\tilde{\bq}}

\newcommand{\sinc}{\mathrm{sinc}}

\newcommand{\vfft}{\tilde{v}}
\newcommand{\vq}{\vel{\bq}}

\newcommand{\gaws}{\gamma_\text{ws}}

\newcommand{\dvw}{\De \vw}

\newcommand{\fluidL}{L_\text{f}}

\newcommand{\tShock}{\ta_\text{sh}}

\newcommand{\TInittil}{\tilde{T}}

\definecolor{newgreen}{RGB}{10,100,20}

\begin{document}

\newcommand{\Sussex}{\affiliation{
Department of Physics and Astronomy,
University of Sussex, Falmer, Brighton BN1 9QH,
U.K.}}

\newcommand{\HIPetc}{\affiliation{
Department of Physics and Helsinki Institute of Physics,
PL 64, % (Gustaf H\"{a}llstr\"{o}min katu 2),
FI-00014 University of Helsinki,
Finland
}}

\title{Sound shell model for acoustic gravitational wave production at a first-order phase transition in the early Universe}
\author{Mark Hindmarsh}
\email{m.b.hindmarsh@sussex.ac.uk}
\Sussex
\HIPetc

\date{\today}

\begin{abstract}
A model for the acoustic production of gravitational waves at a first order phase transition is presented. 
The source of gravitational radiation is the sound waves generated by the explosive growth of bubbles of the stable phase. The model assumes that the sound waves are linear and that their power spectrum is determined by the characteristic form of the sound shell around the expanding bubble.
The predicted power spectrum has two length scales, the average bubble separation and the sound shell width when the bubbles collide. 
The peak of the power spectrum is at wavenumbers set by the sound shell width. For higher wavenumber $k$, the power spectrum decreases as $k^{-3}$. At wavenumbers below the inverse bubble separation, the power spectrum goes as $k^5$.  For bubble wall speeds near the speed of sound where these two length scales are distinguished, there is an intermediate $k^{1}$ power law. The detailed dependence of the power spectrum on the wall speed and the other parameters of the phase transition 
raises the possibility of their constraint or measurement at a future space-based gravitational wave observatory such as eLISA.
\end{abstract}
% \preprint{HIP-2016-XX/TH}
\maketitle

%\section{Introduction}

Interest in gravitational waves from a first order electroweak phase transition in the early Universe \cite{Steinhardt:1981ct,Witten:1984rs,Hogan:1984hx} has greatly increased following 
the European Space Agency's approval of a space-based gravitational wave observatory %, eLISA 
\cite{Seoane:2013qna}
and the detection of gravitational waves from a merging black hole binary \cite{Abbott:2016blz}. At the same time, it has been realised that early work on gravitational waves from a thermal phase transition 
\cite{Kosowsky:1992rz,Huber:2008hg} greatly underestimated their energy density \cite{Hindmarsh:2013xza}.  The first 3-dimensional hydrodynamic simulations  \cite{Hindmarsh:2013xza,Hindmarsh:2015qta} revealed that the dominant source of gravitational waves was acoustic production from sound waves generated by the explosive growth of bubbles of the stable phase. In fact, it had been pointed out long ago that sound waves were a source of gravitational waves \cite{Hogan:1986qda}, but subsequent work had not appreciated that the sound wave source persisted for long after the phase transition completed, hence boosting the signal by orders of magnitude. The original model of gravitational radiation from the colliding bubble walls may still be relevant for near-vacuum transitions \cite{Weir:2016tov}, and a semi-analytic approach has recently been developed \cite{Jinno:2016vai}.

The simulations in \cite{Hindmarsh:2013xza,Hindmarsh:2015qta} revealed a power spectrum peaked at a wavelength around the average bubble separation $\Rbc$, with a power-law $k^{-p}$ at wavenumber $k \gg \Rbc^{-1}$. Where the power law is clear, the index was somewhere in the range $-3 \lesssim p \lesssim -4$.  There was also evidence for some structure in the peak: where the bubble wall speed $\vw$ was closer to the speed of sound, the peak was broader.  Understanding the gravitational wave power spectrum is of great importance for eLISA's detection prospects \cite{Caprini:2015zlo}, and so it is vital to have a better physical understanding of the numerical simulations.

In this work, I outline a model for the acoustic gravitational wave power spectrum based on the observation that the shells of compression and rarefaction (i.e.~sound pulses) around the expanding bubble of the stable phase continue to propagate after the phase boundaries driving them have disappeared. The radial velocity field $v(r,t)$ surrounding the bubble takes a characteristic invariant profile $\vip(\xi)$, with $\xi = r/t$, which acts as the initial condition for the sound wave at the bubble wall collision time $\tInit$. The subsequent local fluid velocity is the superposition of the sound waves from many sound shells, and can be treated as a Gaussian random field while linearity is maintained. The power spectrum of the velocity field is computable from the velocity profile $\vip(\xi)$, which has a simple form in linearised hydrodynamics. The gravitational wave power spectrum can then be computed from a Gaussian velocity field by a simple convolution of the power spectrum \cite{Kosowsky:2001xp,Caprini:2006jb}. 
The model is distinguished from earlier modelling of the velocity field and shear stress correlations  \cite{Caprini:2007xq,Caprini:2009fx} by the recognition that long-lasting sound waves are the main source of gravitational radiation, and by the computation of their power spectrum from the hydrodynamic solution. 

In the sound shell model 
the gravitational wave power spectrum has a characteristic double broken power law form (\ref{e:GWPowLaw}) with two length scales, the bubble separation $\Rbc$ and the sound shell thickness $\De\Rbc$. For wall speeds near the speed of sound $\cs$, the sound shell is thin, and there is a characteristic $k^{1}$ power law in the range  $\Rbc^{-1} \lesssim k \lesssim \De\Rbc^{-1}$. 
For large $k\De\Rbc$, the power law index is $-3$, and for small $k\Rbc$, it is $+5$. 

In the following I recap how the gravitational wave power spectrum can be derived from the shear stress unequal time correlator (UETC), and how the shear stress power spectrum is found from the velocity power spectrum.  I then present the model for the velocity power spectrum, derived from an incoherent superposition of the power spectra from randomly placed sound shells, which are launched into free propagation on the collision of the phase boundaries.

In its simplest form, the model assumes that all fluid velocities are non-relativistic, and that the bubble separation is much less than the Hubble distance at the transition $1/\Hc$. A corollary is that the duration of the transition is much shorter than the Hubble time.
We find it convenient to distinguish between the spectral density of a field with Fourier coefficients $f_\bk$,
$\SpecDen{f} = |f_\bk|^2$, and the power spectrum  
$\Pspec{f} = {k^3}|f_\bk|^2/{2\pi^2}$.

%\section{Gravitational waves from fluid}

We start by defining 
\ben
\tau^{\ij} = w \ep v^iv^j
\een
where $v^i$ is the fluid velocity, $\ep$ is the energy density and $w$ is the equation of state of parameter of the fluid (equal to 1/3 for an ideal relativistic gas).  This is the relevant part of the energy-momentum tensor for gravitational wave production from a fluid moving non-relativistically.

The fluid shear stress UETC $\Pi^2$ is then defined from 
\ben
\la_{ij,kl}(\bk_1)\vev{\ta^{ij}_{\bk_1}(t_1)\ta^{kl}_{\bk_2}(t_2)} = \uetcTen(k_1,t_1,t_2) \debar3{\bk_1 + \bk_2},
\een
where $\la_{ij,kl}(\bk)$ is the transverse-traceless projector for symmetric tensors. 
Assuming that the fluid shear stress fluctuations have a characteristic length scale $\fluidL$ and are stationary well after they are created, we may write
\ben
\Pi^2(k,t_1,t_2) \simeq (w\bar\ep\fluidV^2)^2\fluidL^3\tilde\Pi^2(k\fluidL,\ze),
\een
where $\ze = k(t_1 - t_2)$, $\bar\ep$ is the mean energy density, 
and $\fluidV^2$ is the mean square fluid velocity (weighted by energy density). 
We assume that the energy density fluctuations are of order $\fluidV$ and can be neglected.

The gravitational wave power spectrum of an acoustic source with lifetime $\tLife$ and a length scale $\fluidL$ operating when the Hubble rate is $\Hc$ can be shown to be \cite{Hindmarsh:2015qta} 
\ben
\label{e:GWPowSpe}
\frac{d \OmGW(k)}{d \ln(k)} =  3(1+w)^2 \fluidV^4  (\Hc \tLife) (\Hc\Rfluid) \frac{(k\Rfluid)^3}{2\pi^2} \SpecDenGW(k\Rfluid).
\een
The dimensionless spectral density $\SpecDenGW(k\Rfluid)$ is found from integrating the scaled fluid shear stress UETC with the appropriate Greens functions,
\ben
\label{e:SpecDenInt}
\SpecDenGW(k\Rfluid)  = \frac{1}{k\Rfluid}\int d\ze \frac{\cos(\ze)}{2} \tilde\Pi^2(k\Rfluid,\ze).
\een
Although apparently a Minkowski space expression, it was shown in \cite{Hindmarsh:2015qta} that 
(\ref{e:SpecDenInt}) also applies to an expanding universe, provided the correlation scale $\fluidL$ is much less than the Hubble distance.
It was also shown that viscous damping is negligible for the scales of interest, 
and that the effective source lifetime $\tLife$ is precisely the Hubble time $\Hc^{-1}$ \cite{Hindmarsh:2015qta}. 
Dissipation can also arise through the formation of shocks (and eventually turbulence), after a time \cite{LanLifFlu,Pen:2015qta}
\ben
\tShock \sim \fluidL/\fluidV.
\een
We will assume that $\fluidV \ll \fluidL \Hc $.

%\section{Fluid shear stress from random velocities}

We now assume that the shear stress UETC is generated from a Gaussian random irrotational velocity field $v^i(\bx,t)$, as appropriate for sound waves.  
We denote the Fourier transform 
\ben
\vfft^i_\bq(t) = \int d^3 x v^i(\bx,t)e^{-i\bq\cdot \bx},
\een
and the longitudinal part of its 
%power spectral density by $\SpecDen{v}$, where 
UETC by 
\ben
G(q,t_1,t_2) = \hat{q}^i \hat{q}^j \vev{\vfft^i_{\bq_1}(t_1)\vfft^{*j}_{\bq_2}(t_2)}\debar3{\bq_1 - \bq_2}.
\een
It can then be shown that \cite{Kosowsky:2001xp,Caprini:2006jb,Caprini:2007xq}
\begin{align}
\Pi^2(k;t_1,t_2) 
&= \nonumber\\
\left( \frac{4}3\bar\ep \right)^2  &    \int \dbar3{q} \frac{q^2}{\qtil^2} (1-\mu^2)^2
G(q,t_1,t_2)G(\qtil,t_1,t_2),
%\Pi^2(k;t_1,t_1)
%4 \cos[\om(t_1-t_2)] \cos[\omtil(t_1-t_2)].
\end{align}
where $\bqtil = \bq - \bk$ and $\mu = \hat{\bq}\cdot\hat{\bk}$.

%\section{Velocity power spectrum from random bubbles}
%\label{s:VelPowSpe}

Our model velocity field is the superposition of velocity fields from bubbles nucleated at random times $t^{(n)}$ and positions $\bx^{(n)}$,
\ben
v_i(\bx,t) = \sum_n v^{(n)}_i(\bx,t).
\een
The velocity field of the bubbles is taken to be the asymptotic invariant profile, which is radial. 
When a segment of a bubble wall collides at $\tInit$, removing the local forcing of the fluid, 
the fluid is launched into free propagation, so that 
the invariant profile is the initial condition for the subsequent linear evolution of the velocity field.
The free velocity field obeys the equation 
\ben
\left(\pa_t^2 - \cs\na^2 \right)v^i(\bx,t) = 0, 
\een
whose general solution has the plane wave decomposition 
\ben
v^i(\bx,t) = \int \dbar3{q}\left( \vq^i e^{-i\om t + i \bq\cdot\bx} + \vq^{*i} e^{i\om t - i \bq\cdot\bx} \right), 
\een
where $\om = \cs q$.
Note the distinction between the plane wave amplitudes $\vq^i$ and the Fourier transform of the velocity field  $\vfft^i_\bq(t)$. 

Writing $\dot{\tilde{v}}_i(\bq,t)$ for the acceleration field, 
we see that at the matching time $\tInit$
the plane wave amplitude is 
\ben
\vel{\bq}^i = \half \left( \tilde{v}^i_\bq(\tInit) + \frac{i}{\om} \dot{\tilde{v}}^i_\bq(\tInit) \right) e^{i \om \tInit}.
\een
The fields $v^i(\bx,t)$ and $\dot v^i(\bx,t)$ are related by the 
%initial condition set by the 
fluid velocity around the colliding bubbles, and so the plane wave coefficients $\vel{\bq_1}^i$ and $\vel{\bq_2}^{*j}$ are not independent. We will see (\ref{e:vfft_A}) that the relationship is
\ben
\label{e:vvstar}
\vel{\bq,i}^{(n)} \simeq - e^{2 i\om\tInit + i\theta(z)} \vel{-\bq,i}^{(n)*},
\een
where $\th$ is a $q$-dependent phase. 
With this is mind, one can show that 
\begin{widetext}
\begin{align}
\label{e:VuetcFull}
\vev{\vfft^i_{\bq_1}(t_1)\vfft^{*j}_{\bq_2}(t_2) } 
=  2\hat{q}_1^i \hat{q}_1^j \SpecDen{v}(q_1)\left\{\cos[\om_1(t_1-t_2)] \debar3{\bq_1 - \bq_2} 
-  \cos[\om_1(t_1+t_2 - 2\tInit - \th)] \debar3{\bq_1 + \bq_2}\right\},
\end{align}
\end{widetext}
where $\SpecDen{v}$ is the spectral density of the plane wave amplitudes, defined from 
\ben
\vev{\vel{\bq_1}^i \vel{\bq_2}^{*i} } = \SpecDen{v}(q_1)\debar3{\bq_1 - \bq_2}.
\een
In the sound shell model, the power spectrum is the incoherent sum of contributions to the power spectrum from collisions at a distribution of times $\tInit$.  We assume that this distribution varies slowly over the period of the sound waves, and so that the second term in (\ref{e:VuetcFull}) does not contribute. Hence 
the relation between the velocity unequal time correlation function and the power spectral density of the plane wave coefficients is 
\ben
G(q,t_1,t_2) = 2 P_v(q)\cos[\om(t_1-t_2)],
\een
where $\om = \cs q$.
We see that after integration over all the collisions, the velocity UETC, and therefore the shear stress UETC,  are stationary (depend only on $t_1-t_2$).

With this form of the velocity UETC, 
the cosines combine to produce delta-functions $\de(k\pm\om\pm\omtil)$ in the integral over $t_1 - t_2$
in (\ref{e:SpecDenInt}). 
Only $k-\om - \omtil$ can vanish, and it follows that 
\begin{widetext}
\ben
\SpecDenGW(y) = \frac{1}{4\pi y\cs} \left( \frac{1 - \cs^2}{\cs^2}\right)^2 \int_{z_-}^{z_+} \frac{dz}{z}    \frac{(z - z_+)^2(z-z_-)^2}{(z_+ + z_- - z)} \bar{\SpecDen{v}}(z)\bar{\SpecDen{v}}(z_+ + z_- - z),
\een
\end{widetext}
where $y = k\fluidL$, $z=q\fluidL$ and $z_\pm = \frac{y}{\cs}\left( \frac{1 \pm \cs}{2}\right)$.

It is then clear that if the velocity power spectrum $\Pspec{v}$ goes as $(q\fluidL)^{n}$ over a range of wavenumbers, the gravitational wave power spectrum goes as 
\ben
\frac{d \OmGW(k)}{d \ln(k)} \sim (k\fluidL)^{2n-1}.
\een

We now calculate the spectral density of the plane wave coefficients for the sound shell around a single bubble, nucleated at time $t^{(n)}$ at position $\bx^{(n)}$. The velocity field is radial
%, and approaches a function of radius divided by the time since nucleation, 
and self-similar, 
so that we can write
\ben
%v^{(n)}_i(\bx,t) = \frac{R^{(n)}_i}{R^{(n)}} \hat{R}^{(n)}_i\vip(\xi),
v^{(n)}_i(\bx,t) = \hat{R}^{(n)}_i\vip(\xi),
\een
where $R_i^{(n)} = x_i - x^{(n)}_i$, $T^{(n)} = t - t^{(n)}$, and $\xi = R^{(n)}/T^{(n)}$. 
The function $\vip$ for various wall speeds can be seen in Fig.\ 3 of Ref.~\cite{KurkiSuonio:1995pp}. 
The width of velocity profile is determined by $\De\vw = \vw-\cs$. 

The Fourier transform of the velocity field is 
\ben
\tilde{v}^{(n)}_i(\bq,t)   = e^{-i\bq\cdot\bx^{(n)}} (T^{(n)})^3  i \hat{z}^i f'(z),  
\een
where $z^i = q^iT^{(n)}$ and 
\ben
\label{e:fSinTra}
f(z) = \frac{4\pi}{z} \int_0^\infty d\xi \vip(\xi) \sin(z\xi).
\een
The time derivative of $\vfft^i$ follows from the time-dependence of the invariant profile, so that
\ben
\dot {\tilde{v}}^{(n)}_i(\bq,t)   = e^{-i\bq\cdot\bx^{(n)}}  (T^{(n)})^2 i \hat{z}_i g'(z),  
\een
where 
\bea
g(z) &=& \frac{1}{z} \frac{d}{dz} (z^2 f(z)). 
\eea
The contribution from the $n$th bubble to the coefficient in the plane wave expansion of the general solution is therefore
\ben
\label{e:vfft_A}
\vel{\bq,i}^{(n)} = i (\TInit^{(n)})^3 \hat{z}_i e^{i\om\tInit - i\bq\cdot\bx^{(n)}} A(z), 
\een
where $\TInit^{(n)} = \tInit - t^{(n)}$ and 
\ben
A(z) = \half\left[f'(z) + \frac{i}{\cs z} g'(z)\right].
\een
Note that now,  $z = q\TInit^{(n)}$.

We average over bubble centres and the collision time distribution $n(\TInit)$, which is calculable from the bubble nucleation rate \cite{Guth:1981uk,Enqvist:1991xw}. 
On dimensional grounds we can write
\ben
n(\TInit) d\TInit = \frac{\be}{\Rbc^3} \nu(\be \TInit){d\TInit},
\een
where $\be = (8\pi)^{\frac13} \vw/\Rbc$ \cite{Enqvist:1991xw} is 
the phase transition rate parameter.  By definition, $\int n(\TInit) d\TInit = 1/\Rbc^3$, the bubble number density.
Hence the velocity spectral density is
\ben
\SpecDen{v}(q) =  \frac{\Rbc^3}{(8\pi)^2\vw^6} \int {d\TInittil} \nu(\TInittil) \TInittil^6  |A(\TInittil q/\be)|^2 ,
\een
where $\TInittil = \be\TInit$.
The collision times are peaked around $\TInittil \simeq \be\Rbc/\vw$ with similar width, so if there is a power law $z^{n-3}$ in the 1-bubble spectral density $|A^2|$, we have 
\ben
\SpecDen{v}(q) \sim {\Rbc^3} (q\Rbc)^{n-3}. 
\een
Oscillatory features at high $q\Rbc$ will be averaged over.

The linearised fluid equations have the solution 
\cite{Kamionkowski:1993fg,Huber:2013kj} 
\ben
\label{e:Vip}
\vip(\xi) = \vmax \frac{\vw^2}{\xi^2}\frac{\cs^2- \xi^2}{\cs^2-\vw^2},
\een
where 
$\vmax(\vw,\al)$ is computable %from the energy-momentum conservation equations at the phase boundary 
from the wall speed and the scalar potential difference relative to the total energy density $\al$  \cite{LanLifFlu,Steinhardt:1981ct,Espinosa:2010hh}.
For small $\al$, and $|\vw - \cs| > O(\sqrt{\al}\cs)$, 
\begin{equation}
\vmax \simeq 3\al\vw|\gaws^2|
\end{equation}
where $\gaws^2 = \cs^2/(\cs^2 - \vw^2)$. 
For low fluid speeds, the solution (\ref{e:Vip}) is valid for $\xi$ between $\vw$ and $\cs$. 

The transform (\ref{e:fSinTra}) can be calculated exactly,
giving
\begin{align}
f'(z) &= %\nonumber\\
-4\pi \vmax \vw^2 \cs  \frac{1}{\cs z} \bigg[ \sinc(\vw z) 
\nonumber\\
& +  \frac{2\gaws^2}{(\cs z)^2} \left(\cos(\cs z) - \cos(\vw z)\right)
\bigg],
\\
g'(z) &= %\nonumber\\
4\pi \vmax \vw^2 \cs  \frac{1}{\cs z} \bigg[ \vw z j_1(\vw z) 
\nonumber\\
& +  {2\gaws^2} \left(\sinc(\cs z) - \sinc(\vw z)\right)
\bigg].
\end{align}
The limits of the power spectrum can now be extracted, distinguishing between 
the scales $\vw z$ and $\dvw z$.
First, when both $z\vw$ and $z\dvw$ are small,
\begin{align}
|A(z)|^2 & \simeq    \frac{4\pi^2}{9} \vmax^2 {\vw^4 \cs^2}\left(1+\frac{\vw}{\cs}\right)^2\left(\frac{\dvw}{\cs}\right)^2,
\end{align}
a white noise power spectrum as expected.
For a thin sound shell, we can investigate the range 
$z\dvw \ll 1 \ll z\vw$, for which 
\begin{align}
|A(z)|^2 & \simeq  \pi^2\vmax^2 \cs^6\left(\frac{\dvw}{\cs}\right)^2 \frac{1}{(\cs z)^2}.
\end{align}
Finally, at when both $z\vw$ and $z\dvw$ are large,
\begin{align}
|A(z)|^2 & \simeq  4\pi^2\vmax^2 \vw^2\cs^2 \frac{\cs^2\sin^2(\vw z) + \vw^2\cos^2(\vw z)}{(z\cs)^4}.
\end{align}
Hence the velocity power spectrum goes as  
\ben
\label{e:VPowLaw}
\Pspec{v}(q) \sim \left\{ 
\ba{ll}
(q\Rbc)^{3}, &  q\De\Rbc, q\Rbc \ll 1 \\
(q\Rbc)^{1}, & q\De\Rbc \ll 1 \ll q\Rbc \\
(q\Rbc)^{-1}, & 1 \ll q\De\Rbc, \Rbc \\
\ea
\right.
\een
where $\De\Rbc \simeq \De\vw/\be$.
The $q^{-1}$ form at large $q$ is a consequence of the compact support of the velocity field for $\xi$ between $\vw$ and $\cs$. If the discontinuity at $\xi=\vw$ is smoothed over a length scale $\ell$, the power law form no longer applies for $q \gg \ell^{-1}$.

The gravitational wave power spectrum %from (\ref{e:VPowLaw}) 
goes as 
\ben
\label{e:GWPowLaw}
\frac{d \OmGW(k)}{d \ln(k)} \sim \left\{ 
\ba{ll}
(k\Rbc)^{5}, &  k\De\Rbc, k\Rbc \ll 1, \\
(k\Rbc)^{1}, & k\De\Rbc \ll 1 \ll k\Rbc, \\
(k\Rbc)^{-3}, & 1 \ll k\De\Rbc, \Rbc. \\
\ea
\right.
\een
The form (\ref{e:GWPowLaw}) is the main result.
Note that the peak of the power spectrum is set by the sound shell thickness, not the bubble separation.
Note also that the low wavenumber power law is steeper than the expected $k^3$. This is consistent with causality, which only bounds the power spectrum to be less than $k^3$. At very low wavelengths, we expect the contribution from bubble collisions, which does go as $k^3$, to take over.

We can check these predictions against the velocity and gravitational wave power spectra from numerical simulations, shown in Figs.~7 and 8 of Ref.~\cite{Hindmarsh:2015qta}. 
As noted in that paper, at high $k$ one can identify a $k^{-1}$ power law in the velocity power spectrum and a $k^{-3}$ power law in the gravitational wave power spectrum for the weak deflagration ($\vw = 0.44$, $\fluidV \simeq 0.0073$), in accord with the sound shell model. The power spectra are steeper for the weak detonation ($\vw = 0.83$, $\fluidV \simeq 0.0052$), although this may be due to the velocity field not reaching its invariant profile (see Fig.~3), or a lack of dynamic range beyond the peak. Interestingly, at $\vw = 0.54$ there is a suggestion of an intermediate $k^{1}$ power law in both the velocity and gravitational wave power spectra, although there is not enough dynamic range to clearly distinguish them, or a long-wavelength $k^5$ power law in gravitational waves. 
Further tests require larger simulations \cite{Hin+16}. 

%\section{Discussion and conclusions}

In summary, a model for the acoustic production of gravitational waves at a first-order phase transition has been outlined. 
The gravitational wave power spectrum is a double broken power law (\ref{e:GWPowLaw}) built from the two physical scales: the bubble separation and the sound shell thickness.  For a generic wall speed these scales are approximately the same, and the form of the power spectrum for both deflagrations and detonations is similar (but not identical). Wall speeds near the speed of sound are distinguished by an intermediate $k^{1}$ power law between the two scales. 

The detailed power spectrum depends computably on the wall speed, the bubble separation, the Hubble parameter at the transition, and the latent heat. 
Future work will exhibit this dependence, and explore prospects for the measurement or constraint of the phase transition parameters by accurate determination of a stochastic gravitational wave background. 

I am grateful to Nicola Hopkins for collaboration on early stages of this work, and to Stephan Huber, Kari Rummukainen, David Weir for many discussions. 
I have also benefited from exchanges with Chiara Caprini, Germano Nardini and other members of the eLISA Cosmology Working Group. 
I acknowledge support from the Science and Technology Facilities Council (grant number ST/J000477/1).

\bibliography{GWs}

\end{document}